%% file: main.tex
\documentclass[sigconf]{acmart}
\AtBeginDocument{%
  }

\copyrightyear{2025}
\acmYear{2025}
\setcopyright{cc}
\setcctype{by}
\acmConference[SC Workshops '25]{Workshops of the International Conference
for High Performance Computing, Networking, Storage and Analysis}{November
16--21, 2025}{St Louis, MO, USA}
\acmBooktitle{Workshops of the International Conference for High Performance
Computing, Networking, Storage and Analysis (SC Workshops '25), November
16--21, 2025, St Louis, MO, USA}
\acmDOI{10.1145/3731599.3767582}
\acmISBN{979-8-4007-1871-7/2025/11}

\input{preamble}

\begin{document}

\title[LLM Agents for Interactive Workflow Provenance]{LLM Agents for Interactive Workflow Provenance:\\ Reference Architecture and Evaluation Methodology}

\input{_authors}

\thanks{
This is a preprint for a publication in the proceedings of the Supercomputing Conference. Please cite using the ACM Reference Format (on the top-right).\\
{\tiny
Notice: This manuscript has been authored in part by UT-Battelle, LLC under Contract No. DE-AC05-00OR22725 with the U.S. Department of Energy. The United States Government retains and the publisher, by accepting the article for publication, acknowledges that the United States Government retains a non-exclusive, paid-up, irrevocable, world-wide license to publish or reproduce the published form of this manuscript, or allow others to do so, for United States Government purposes. The Department of Energy will provide public access to these results of federally sponsored research in accordance with the DOE Public Access Plan (http://energy.gov/downloads/doe-public-access-plan).
}
}

\renewcommand{\shortauthors}{R. Souza, et al.}

\input{sections/abstract}



\maketitle

\input{sections/intro}
\input{sections/background}
\input{sections/method}
\input{sections/framework}

\input{sections/evaluation}
\input{sections/conclusion}

\input{sections/ack}

\bibliographystyle{ACM-Reference-Format}
\bibliography{references}

\end{document}

%% file: preamble.tex
\usepackage{listings} 
\usepackage{soul}
\usepackage{caption}
\usepackage{graphicx}
\usepackage{enumitem}
\usepackage{wrapfig} 
\usepackage{colortbl}
\usepackage{algorithmic}
\usepackage[utf8]{inputenc}
\usepackage{booktabs}
\usepackage{pifont}  %

\definecolor{graytext}{gray}{0.5}
\definecolor{essential}{RGB}{0,0,0}       
\definecolor{auxiliary}{gray}{0.45}       

\lstdefinelanguage{json}{
    morestring=[b]",
    morecomment=[l]{//},
    morekeywords={true,false,null},
    sensitive=false,
}

\lstdefinestyle{jsonstyle}{
    language=json,
    basicstyle=\tiny\ttfamily, 
    showstringspaces=false,
    backgroundcolor=\color{white},
    commentstyle=\color{graytext},
    keywordstyle=\color{black},
    breaklines=true,
    frame=single,
    numbers=none,
    captionpos=b,
    escapeinside={(*@}{@*)},
    aboveskip=0.5em,
    belowskip=0.5em,
    lineskip=0pt
}

%% file: _authors.tex
\settopmatter{authorsperrow=4} 

\author[R. Souza]{Renan Souza}
\affiliation{%
  \institution{Oak Ridge National
Lab.}
  \city{Oak Ridge}
  \state{TN}
  \country{USA}
}
\email{souzar@ornl.gov}

\author[T. Poteet]{Timothy Poteet}
\affiliation{%
\institution{Oak Ridge National
Lab.}
\city{Oak Ridge}
  \state{TN}
  \country{USA}
}
\email{poteetts@ornl.gov}

\author[B. Etz]{Brian Etz}
\affiliation{%
  \institution{Oak Ridge National
Lab.}
  \city{Oak Ridge}
  \state{TN}
  \country{USA}
}
\email{etzbd@ornl.gov}

\author[D. Rosendo]{Daniel Rosendo}
\affiliation{%
  \institution{Oak Ridge National
Lab.}
  \city{Oak Ridge}
  \state{TN}
  \country{USA}
}
\email{rosendod@ornl.gov}

\author[A. Gueroudji]{Amal Gueroudji}
\affiliation{%
  \institution{Argonne National Lab.}
  \city{Lemont}
  \state{IL}
  \country{USA}
}
\email{agueroudji@anl.gov}

\author[W. Shin]{Woong Shin}
\affiliation{%
  \institution{Oak Ridge National
Lab.}
  \city{Oak Ridge}
  \state{TN}
  \country{USA}
}
\email{shinw@ornl.gov}

\author[P. Balaprakash]{Prasanna Balaprakash}
\affiliation{%
  \institution{Oak Ridge National
Lab.}
  \city{Oak Ridge}
  \state{TN}
  \country{USA}
}
\email{pbalapra@ornl.gov}

\author[R. Ferreira da Silva]{Rafael Ferreira da Silva}
\affiliation{%
  \institution{Oak Ridge National
Lab.}
  \city{Oak Ridge}
  \state{TN}
  \country{USA}
}
\email{silvarf@ornl.gov}

%% file: sections/abstract.tex
\begin{abstract}
Modern scientific discovery increasingly relies on workflows that process data across the Edge, Cloud, and High Performance Computing (HPC) continuum. Comprehensive and in-depth analyses of these data are critical for hypothesis validation, anomaly detection, reproducibility, and impactful findings. Although workflow provenance techniques support such analyses, at large scale, the provenance data become complex and difficult to analyze. Existing systems depend on custom scripts, structured queries, or static dashboards, limiting data interaction. In this work, we introduce an evaluation methodology, reference architecture, and open-source implementation that leverages interactive Large Language Model (LLM) agents for runtime data analysis. Our approach uses a lightweight, metadata-driven design that translates natural language into structured provenance queries. Evaluations across LLaMA, GPT, Gemini, and Claude, covering diverse query classes and a real-world chemistry workflow, show that modular design, prompt tuning, and Retrieval-Augmented Generation (RAG) enable accurate and insightful LLM agent responses beyond recorded provenance.
\end{abstract}

%% file: sections/intro.tex
\section{Introduction}

Modern scientific discovery depends on workflows that process multimodal data across the Edge, Cloud, and HPC continuum (ECH)~\cite{antypas2021enabling}. These workflows are undergoing a transformative shift with the emergence of autonomous agents powered by LLMs or other foundation models, capable of planning, making decisions, and coordinating interactions with humans, other agents, and running tasks~\cite{silva2025wisdom, pauloski2025empowering}. Comprehensive and in-depth data analyses in this context are paramount for hypothesis validation, anomaly detection, reproducibility, and ultimately new findings that impact society. However, such analyses are challenging due to the highly distributed and heterogeneous infrastructures, data, and systems. 

Workflow provenance has long supported data analyses through techniques that track data and tasks~\cite{mattoso_scientific_2010}.
Nevertheless, as workflows grow in complexity and scale, provenance data become increasingly intricate and difficult to analyze.
Existing approaches~\cite{souza2020workflow} rely on custom scripts, structured queries, or fixed dashboards, limiting interactivity and flexibility for exploratory analysis and distancing scientists from the data-to-insights process.

This paper introduces a modular, loosely coupled provenance agent system architecture that facilitates live interaction between users and data during workflow execution in the ECH continuum. The architecture leverages LLM-powered agents and the Model Context Protocol (MCP)~\cite{mcp} to support natural language interactivity.
It adopts a lightweight, metadata-driven design that translates natural language into accurate runtime workflow provenance queries. Specifically, we make the following contributions:

\begin{itemize}[leftmargin=*,nosep]

    \item \textbf{Methodology for evaluating LLMs in workflow provenance interaction.} We present a domain-agnostic, system-independent methodology focused on RAG pipeline design and prompt engineering to assess LLM performance on diverse workflow provenance \textit{what}, \textit{when}, \textit{who}, and \textit{how} query classes, enabling extensibility for other systems and research.

    \item \textbf{Reference architecture for a provenance-aware AI agent framework for live interaction and data monitoring}. The architecture enables LLM-driven interactions such as natural language querying and data visualization. It enforces separation of concerns across context management, prompt generation, LLM services, tool dispatching, and provenance data access via database and streaming layers.

    \item \textbf{Open-source implementation of the agent framework atop a flexible, broker-based provenance infrastructure.} The system~\cite{flowcept} builds on a loosely-coupled architecture for distributed provenance capture that supports instrumented code and data observability across ECH workflows. While this architecture is not the focus of this paper, it provides a foundation for deploying and evaluating the provenance-aware agent.

    \item \textbf{Evaluation of LLaMA, GPT, Gemini, and Claude LLMs on diverse provenance query classes, and demonstration of the agent in a real-world computational chemistry workflow.} Leveraging a lightweight, dataflow schema-driven RAG pipeline with guidelines, the agent scales with workflow complexity and generalizes to other workflows with accurate, interactive, runtime data exploration without domain-specific tuning.

\end{itemize}

%% file: sections/background.tex
\section{Background and Related Work}

\subsection{Data Analysis via Workflow Provenance}

\textit{Data provenance}, also known as data lineage, refers to metadata that records how data was generated. \textit{Workflow provenance} expands this notion to describe the structure, control logic, and execution details of computational workflows. \textit{Workflow provenance data} encompasses the runtime metadata emitted as workflows execute, capturing what was done, when, where, how, and by whom~\cite{mattoso_scientific_2010}. 
\textit{Workflow task provenance} refers to data captured during the execution of a workflow task. In traditional workflow engines~\cite{suter2025fgcs}, this typically corresponds to a scheduled and executed task. In the context of scripts, it may represent a function call or a logical, cohesive block of code.
In computational science, such metadata are essential for critical capabilities such as reproducibility, hypothesis validation, transparency, explainability, anomaly diagnosis, especially in complex workflows that span the ECH continuum~\cite{prov_ech_continuum}. 

\sloppy
Provenance data encompass multiple semantic dimensions: dataflow, which describes how inputs and outputs are connected and transformed across tasks; control flow, which contain task dependencies and execution order; telemetry, which includes performance metrics such as CPU and GPU usage, memory consumption, and execution times; and scheduling, which identifies where tasks were executed, including hardware placement~\cite{souza2020workflow}.

Query workloads over this metadata vary in scope and behavior. Some involve targeted queries, which filter specific tasks or fields, while others require graph traversal to analyze multi-step dependencies or causal chains. These queries may follow online analytical processing (OLAP) patterns for exploration and monitoring or online transactional processing (OLTP) patterns for fast and targeted lookups.
Provenance can also be classified by its nature: retrospective provenance records actual workflow execution, while prospective provenance defines planned workflow structure~\cite{freire2008}. 

\begin{figure}[!ht]
  \centering
  \includegraphics[width=\columnwidth]{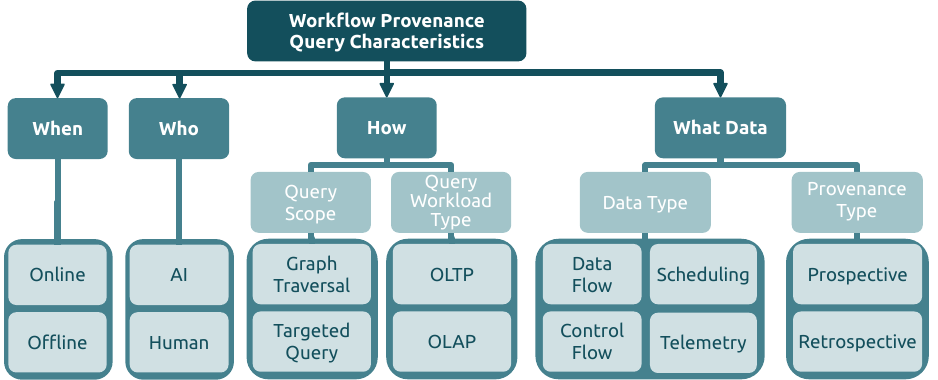}
  \caption{Taxonomy of workflow provenance query characteristics used to define query classes.}
  \label{fig:query_classes}
\end{figure}

Finally, provenance analysis can occur offline, after workflow completion, or online, during execution. In both modes, human users or AI agents may act as producers and consumers of provenance data. Figure~\ref{fig:query_classes} summarizes the provenance characteristics considered in this work, with leaf nodes representing the query classes used in our methodology.

\subsection{LLMs and AI Agents}

LLMs are deep neural networks trained on massive text corpora to perform tasks like question answering and code generation. When integrated with tools, memory, or user interaction loops, they become agents capable of reasoning. These agents operate via prompts, i.e., structured text inputs that guide behavior, whose quality strongly influences outcomes. Prompt engineering involves techniques like instruction tuning and few-shot examples, which can be enhanced by RAG dynamically injecting external knowledge, e.g., from databases or webpages~\cite{llm_prompt}.
Since LLMs process inputs and outputs as tokens, the prompt length must stay within the model's context window, which varies by model. Parameters like temperature control output variability~\cite{llm_prompt}.
MCP~\cite{mcp} is emerging as a standard for AI application development, integrating LLMs with external systems. It defines key concepts such as tools, prompts, resources, context management, and agent–client architecture.


\subsection{Reference Architecture for Distributed Workflow Provenance}
\label{sec:ref-architecture}

\begin{figure}[!ht]
  \centering
  \includegraphics[width=\columnwidth]{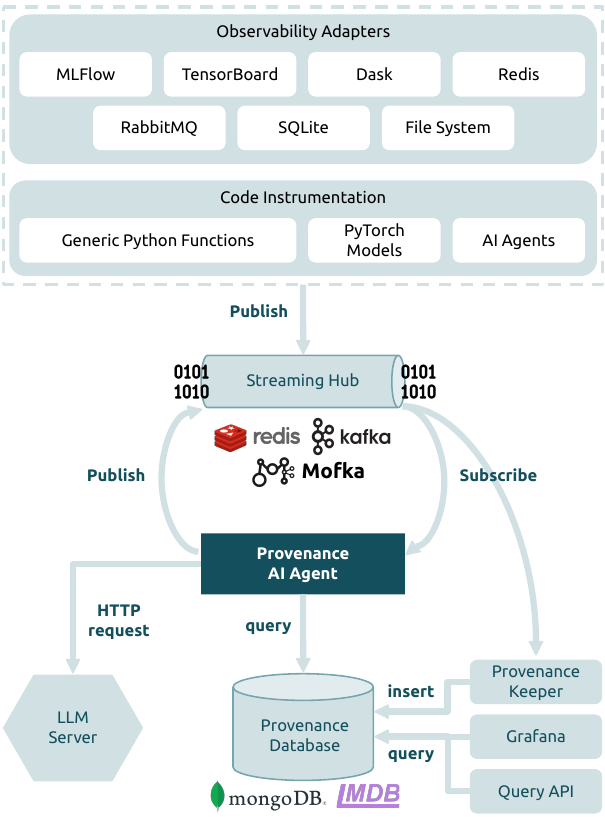}
  \caption{Reference architecture for distributed workflow provenance with a provenance AI agent.}
  \label{fig:prov-architecture}
\end{figure}

Rather than designing a new provenance system from scratch, we build on a reference architecture (Figure~\ref{fig:prov-architecture}) that follows established distributed design principles and is extensible to support an LLM-based provenance agent. This approach provides a flexible, loosely coupled framework that scales from small centralized workflows to large HPC workflows across the ECH continuum.

The architecture prioritizes interoperability and deployment flexibility~\cite{souza2023flowcept} through a modular, adapter-based design that supports provenance capture via two complementary mechanisms: (i) non-intrusive observability adapters, which passively monitor dataflow from services such as RabbitMQ, SQLite, MLflow, and file systems without modifying application code; and (ii) direct code instrumentation, which uses lightweight hooks such as Python decorators to capture fine-grained task-level metadata from functions, scripts, or packages like PyTorch and MCP SDK,  providing flexible entry points for capturing provenance across heterogeneous workflows. To reduce interference with HPC applications, provenance messages are buffered in-memory and streamed asynchronously to a central hub using a publish-subscribe protocol with configurable flushing strategies. Listing~\ref{lst:task-object} shows an example of such a message.

\input{codes/prov_task}

Each component can be independently deployed across the ECH continuum, as long as they can share access to the streaming hub. For lightweight deployments, a single broker may suffice, while large-scale ECH workflows can benefit from federated hubs composed of multiple brokers tailored to specific performance and reliability needs. 
For example, Redis offers low-latency messaging with minimal setup, making it suitable for most use cases; Kafka enables high throughput streaming for data-intensive workflows; and Mofka provides RDMA-optimized transport ideal for tightly coupled HPC networks~\cite{mofka}. Regardless of the underlying broker, all provenance messages adhere to a common schema.

One or more distributed Provenance Keeper services subscribe to the streaming hub, convert incoming messages into a unified workflow provenance schema based on a W3C PROV extension~\cite{souza2025provagent}, and store them in a backend-agnostic provenance database.
The architecture is designed to support multiple DBMS options, including MongoDB for filtering and aggregation, LMDB for high-frequency key–value inserts, and Neo4j for graph traversal queries. 
Users can access provenance data through a language-agnostic Query API, either programmatically (e.g., via Jupyter), through dashboards such as Grafana, or, as introduced in this work, via natural language.

\input{sections/related_work}

%% file: codes/prov_task.tex
\begin{figure}[!ht]
  \noindent\begin{minipage}{\linewidth}
    \begin{lstlisting}[style=jsonstyle, 
      caption={Example of a workflow task provenance message from the chemistry workflow used in this work.},
      label={lst:task-object}]
{
  "task_id": "1753457858.952133_0_3_973",
  "campaign_id": "0552ae57-1273-4ef8-a23b-c5ae6dd0c080",
  "workflow_id": "4f2051b9-cfa3-4ef5-b632-907a3be06899",
  "activity_id": "run_individual_bde",
  "used": {
    "e0": -155.033799510504,
    "frags": {
      "label": "C-H_3",
      "fragment1": "[H]OC([H])([H])[C]([H])[H]",
      "fragment2": "[H]"
    },
    "h0": 0.08547606488512516,
    "outdir": "bde_calc",
    "s0": 0.064344,
    "z0": 0.08026498424723788
  },
  "generated": {
    "bond_id": "C-H_3",
    "bd_energy": 98.64865792890485,
    "bd_enthalpy": 100.22765792890056,
    "bd_free_energy": 92.39108332890055
  },
  "started_at": 1753457858.952133,
  "ended_at": 1753457859.009404,
  "hostname": "frontier00084.frontier.olcf.ornl.gov",
  "telemetry_at_start": {"cpu": ["percent": 23.4]},
  "telemetry_at_end": {"cpu": ["percent": 53.8]},
  "status": "FINISHED",
  "type": "task"
}
    \end{lstlisting}
  \end{minipage}
\end{figure}

%% file: sections/related_work.tex
\subsection{Related Work}

Related work is orthogonal and complementary to ours, exploring different intersections of provenance, workflows, and LLMs. 
PROLIT~\cite{gregori2025prolit}, the most closely related, uses LLMs to rewrite data pipelines for provenance capture, focusing on enhancing completeness and semantics, differing from our approach that focuses on a general-purpose agent architecture for runtime provenance capture and live interaction across ECH workflows. 
Other efforts use LLMs to semantically enrich system-level provenance for cyberattack detection~\cite{zuo2025llmprovenance}. TableVault~\cite{zhao2025tablevault} captures in LLM-augmented workflows but does not address interactive queries. 
Hoque et al.~\cite{10.1145/3613904.3641895} show how LLM provenance capture can document the use of language models in aligning with policies and supporting AI-assisted writing.
LLM4Workflow~\cite{xu2024llm4workflow} focuses on workflow generation, not provenance. Finally, SWARM~\cite{balaprakash2025swarm} and Academy~\cite{pauloski2025empowering} explore intelligent agents for distributed workflows but do not target provenance queries. To our knowledge, this is the first work to define a reference architecture and evaluation method for an LLM-powered agent that supports interactive, schema-aware querying of live workflow provenance data in distributed scientific computing.

%% file: sections/method.tex
\section{LLM-Powered Provenance Agent Evaluation Methodology}
\label{sec:taxonomy-methodology}

\begin{figure*}[!ht]
  \centering
  \includegraphics[width=\textwidth]{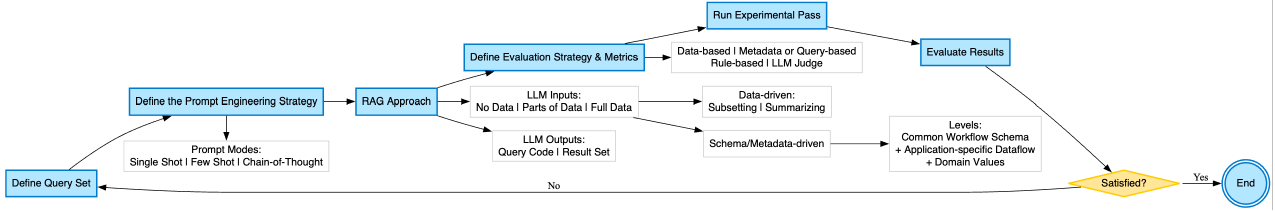}
  \caption{Evaluation methodology to iteratively improve the agent’s performance via prompt and context tuning.}
  \label{fig:methodology}
\end{figure*}

The core function of the LLM-powered provenance agent is to bridge the gap between users and complex provenance databases 
by interpreting natural language queries and producing accurate, context-aware responses.
This requires navigating large and heterogeneous provenance graphs, reasoning over metadata, and in some cases, inferring information not explicitly recorded.
%
To evaluate this capability, we introduce a flexible and extensible methodology designed to assess agent performance across a wide range of provenance query classes and scientific domains. Our methodology guides the development of effective prompt engineering and RAG approaches, helping researchers and developers build agents that are both accurate and generalizable. Unlike prior work that focuses on querying general-purpose databases~\cite{NEURIPS2023_83fc8fab}, our method addresses the specific challenges of workflow provenance, such as domain-dependent dataflow schemas, variable task granularity, and the need to understand the core semantic structures defined by the W3C PROV model~\cite{W3CPROV}, including entities, activities, and agents. This provenance-specific perspective ensures a more targeted evaluation aligned with the goals of scientific data analysis.

Our approach emphasizes the reuse of existing LLMs, including open-source models, to avoid the high cost and complexity of fine-tuning or building custom models. Since LLM performance depends heavily on context, our methodology focuses on designing efficient prompts and RAG pipelines that enrich the agent's input context with contextual metadata, dataflow schema fragments, and representative data.
Figure~\ref{fig:methodology} outlines the six main stages of this process (blue nodes), each with configurable options (white nodes) that support adaptation to different domains and agent designs.

\vspace{3pt}
\noindent \textbf{Query Set.}  
The evaluation begins by defining a \texttt{Query Set}, which serves as the golden dataset. Guided by the query class taxonomy (Figure~\ref{fig:query_classes}), this set enables developers to build a balanced query set composed of natural language queries, their corresponding class labels (taxonomy leaves), and expected answers. When possible, responses should be curated by humans to ensure accuracy.

\vspace{3pt}
\noindent \textbf{Prompt Engineering.}
Defining how queries are formulated, prompt engineering techniques range from \textit{Single-shot}, where only the raw user query is given, to \textit{Few-shot}, which includes additional examples, and \textit{Chain-of-Thought (CoT)}, which extends few-shot prompts with intermediate reasoning steps to support complex queries. Query guidelines can be added to improve accuracy by adding domain-specific instructions.
Yet, the prompt may be enhanced by clear role-giving (e.g., ``you are a workflow provenance specialist'') and job (e.g., ``your job is to interpret the user query and provide a structured query'').

\vspace{3pt}
\noindent \textbf{RAG Strategies.}  
A common RAG use case is accessing data from an external database; here, the raw provenance data. Strategies include: \textit{No data} (no provenance in the prompt), \textit{Partial data} (a relevant subset), and \textit{Full data} (entire database). While full data would be ideal, incorporating large provenance datasets into LLM prompts, directly or via vector-based retrieval, quickly exceeds context limits, even for advanced models~\cite{app14052074}. This challenge grows with workflow size, and even RAG pipelines like LangChain QA chains or summarization approaches struggle to scale.

An alternative is to follow lightweight approaches based on schema or metadata instead of full data. Practical strategies to augment the context are: \textit{common workflow schema}, which adds descriptions for domain-agnostic common fields (e.g., \texttt{task\_id}, \texttt{activity\_id}); \textit{application-specific dataflow fields}, which include domain-specific parameters and outputs; and \textit{semantic descriptions and domain-specific values}, which expand the context with field descriptions and representative example values (which refers to the \textit{partial data} strategy). For instance, in Earth's climate workflows, adding sample values like 0–110 for a field named ``temperature'' can help inferring  range of plausible values or the likely unit (e.g., Fahrenheit), even if the user does not specify it.

Similarly to designing effective input prompts, one needs to carefully approach how the LLM should produce the outputs, also due to context window limits, which include the output tokens. Possible LLM output formats may be: \textit{a result set}, containing the query result; \textit{a structured query}, with the query code to the provenance database; or other text, including summaries of the result dataset and reasoning. Among these strategies, the one that returns the query consumes fewer tokens. The most lightweight combined strategy is the one that uses no or partial data for the inputs and queries for the outputs, as the LLM performance (accuracy, processing time) does not depend on the provenance data size.

\vspace{3pt}
\noindent \textbf{Evaluation.}  
Depending on the output, the evaluation may focus on: \textit{Query-based evaluation}: analyze the generated query according to its syntax, structure, semantics, and used fields; \textit{Result-based evaluation}: compare result sets against ground truth using, e.g., string similarity metrics; and \textit{Hybrid}: metrics combining both.
Then, evaluation methods include: \textit{Rule-based}: match patterns, projection fields, filters, aggregations, etc.; \textit{LLM-as-a-judge}: use  external LLMs specialized to assess correctness~\cite{gu2024surveyllmasajudge}, which can analyze the generated query or result sets; and \textit{Hybrid}: combine rule-based and LLM-based scoring.
While rule-based scoring is transparent and interpretable, often manually curated by humans, it is difficult to design comprehensively and is prone to edge-case errors. By contrast, LLM-as-a-judge methods are more scalable and easier to implement, enabling nuanced evaluations without the need to explicitly encode every rule. However, they introduce opacity, and because LLMs may hallucinate and produce biased results, human oversight remains important to ensure evaluation accuracy.

\vspace{3pt}
\noindent \textbf{Experimental Runs and Refine.}  
With all configurations and evaluation strategies in place, the agent is run across the full \texttt{Query Set}. Evaluation results are analyzed to guide iterative improvements in prompt design, RAG strategy, and agent architecture. This process continues until performance is satisfactory.

%% file: sections/framework.tex
\section{LLM-powered Provenance Agent for Workflow Data Analysis}
\label{sec:agent-architecture}

This section describes the reference architecture of an AI agent that enables live interaction with large-scale streaming provenance data in ECH workflows. The architecture supports both runtime and post-hoc querying, anomaly detection, and LLM reasoning, all while adhering to a modular design and separation of concerns.

\subsection{System Design Decisions}

The provenance agent adopts the loosely coupled, distributed design from Section~\ref{sec:ref-architecture}, making it suitable for ECH workflows. For HPC workloads, we reduce overhead by deploying agent components on a separate node outside the compute job. Workflow tasks perform lightweight provenance capture by buffering messages (see Listing~\ref{lst:task-object}) that are asynchronously streamed in bulk to the hub, reducing interference with active jobs~\cite{souza2023flowcept}.
The provenance database, which persistently consolidates all messages, is hosted externally to maintain a clear separation between the compute job, agent logic, LLM service interaction (typically through REST APIs), provenance capture, and storage. Adopting MCP ensures interoperability with other MCP-compliant agents and systems.

A key challenge in using LLMs to query provenance databases is the limited context window of the models. 
To address this, our approach is driven by the metadata of the provenance data, focusing on capturing and maintaining a compact and semantically meaningful in-memory structure: the \textit{Dynamic Dataflow Schema}.
Rather than submitting raw provenance records directly to the LLM service, the system automatically maintains a schema that summarizes how data flow between tasks, what parameters and outputs are captured, and how workflows evolve over time. 
We do not require users to define this schema upfront, which would hinder usability, introduce complexity, and reduce generalizability to new applications and domains. Instead, the dynamic dataflow schema is incrementally inferred at runtime from live provenance streams and kept up to date by the agent's \texttt{Context Manager}.
Then, the schema is used within the RAG inference pipeline and allows the LLM service to effectively respond to runtime queries even without having access to the actual provenance database.
This approach is also beneficial in scenarios involving sensitive data, where raw records should not be sent to LLM services, particularly when those services are hosted and managed by external parties.

\subsection{Architecture Overview}

\begin{figure}[!ht]
  \centering
  \includegraphics[width=\columnwidth]{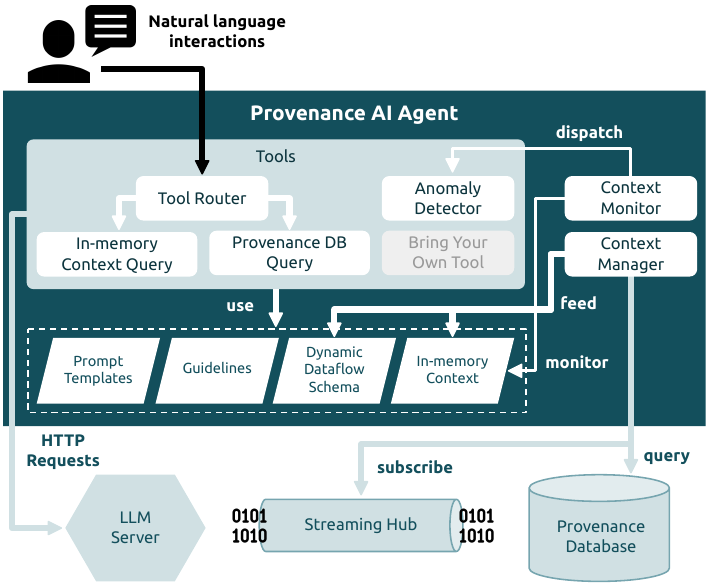}
  \caption{Provenance AI agent architecture design.}
  \label{fig:prov-ai}
\end{figure}

Figure~\ref{fig:prov-ai} presents the architecture of the provenance agent. The architecture is composed of loosely coupled, independently invocable components that communicate asynchronously via a streaming hub service. The major components are described below.

\vspace{3pt}
\noindent \textbf{Context Manager.}
The \texttt{Context Manager} subscribes to the \texttt{Streaming Hub} to receive provenance messages emitted by client workflows. It is responsible for maintaining up-to-date internal in-memory data structures:

\begin{itemize}[leftmargin=*,nosep]
    \item \texttt{In-memory context}, which is a buffer of recent workflow task provenance messages.
  
    \item \texttt{Dynamic Dataflow Schema}, which condenses the dataflow-level semantics of the workflow as it executes, containing each activity and its input and output fields.
    In the task provenance message, the \texttt{used/generated} fields contain the application-specific data captured by the provenance system. For every new incoming raw provenance message, the dataflow schema is updated for these fields, with its name in the message object, inferred data types, and a few example data values for each field. 
    Additionally, the description of fields that are common for all tasks, like \texttt{campaign\_id}, \texttt{workflow\_id}, and \texttt{activity\_id}, is statically included in the schema by default, helping queries that need them.

    \item \texttt{Guidelines}, a dynamic and adaptable set combining domain-agnostic with user-defined instructions that steer the LLM when generating structured queries. These guidelines help resolve ambiguity, enforce preferred conventions (e.g., which field to sort by), and reduce syntax or logic errors. 
    In addition to a static set of domain-agnostic guidelines, users can provide new domain-specific guidelines interactively through natural language (e.g., ``\textit{use the field \texttt{lr} to filter learning rates}''), which are told in the internal prompt to override any other conflicting guideline stated earlier, are stored in the agent's overall context for the current session, and automatically incorporated into future prompts, improving the agent's adaptability and accuracy during multi-turn interactions, especially in domain-specific queries.
\end{itemize}

\vspace{3pt}
\noindent \textbf{Monitoring and Post-hoc Query Tools.}
User-issued natural language queries are handled by a \texttt{Tool Router}, which combines rule-based logic and LLM calls to determine the appropriate handling strategy.
For instance, the LLM response indicates if the user intent is a simple greeting, which does not require any querying, or if the intent is to query the in-memory context (online, monitoring queries) or the persistent database (offline, historical queries).

\vspace{3pt}
\noindent 
\textbf{Context Monitor and Anomaly Detector.}
The \texttt{Context} \texttt{Mon-} \texttt{itor} periodically inspects the in-memory buffer maintained by the \texttt{Context Manager} and dispatches tools based on configurable rules. One such tool is the \texttt{Anomaly Detector}, 
which inspects the data and identifies abnormal patterns in telemetry or domain-specific values (in the \texttt{used} and \texttt{generated} fields), using statistical methods such as outlier detection and standard deviation analysis.
If an anomaly is detected, the system tags the corresponding message with metadata describing the anomaly and publishes a new message to the \texttt{Streaming Hub}. This allows downstream services to detect and react to abnormal tasks, for example, by notifying users. The anomaly tag also enables easier querying and filtering of abnormal tasks. We note that not all MCP tools require LLM interaction, as seen by the \texttt{Anomaly Detector}. Moreover, the architecture is designed to support the addition of new tools, represented with the ``Bring your own tool" box in Figure~\ref{fig:prov-ai}, maintaining the separation of concerns without requiring changes to the core components and taking advantage of the agent’s internal context structures. 

\vspace{3pt}
\noindent 
\textbf{Provenance of Tools and LLM interactions.}
Finally, extending the provenance model, all tool invocations are recorded as workflow tasks, which are subclasses of W3C \texttt{prov:Activity}, with arguments stored as \texttt{prov:used} and results as \texttt{prov:generated}. Each LLM interaction is also stored following the same schema of workflow tasks, but with the prompts filling the \texttt{prov:used} and the LLM response filling the 
\texttt{prov:generated}. If an LLM interaction happened in the context of a tool execution, the tool execution is linked with the LLM interaction via \textit{prov:wasInformedBy}. The agent itself is registered as a \texttt{prov:Agent}, with tool executions and LLM interactions linked to it via \textit{prov:wasAssociatedWith}, enabling traceability of agent-driven analysis.

%% file: sections/evaluation.tex
\section{Experimental Evaluation}

\subsection{Implementation Details and Use Cases}
\label{sec:implementation}

We extend Flowcept (Section~\ref{sec:ref-architecture}) with the provenance agent (Section~\ref{sec:agent-architecture}) using the Python-based MCP SDK. The agent's GUI is implemented using Streamlit, with an accompanying API available for terminal-based or programmatic access. Provenance messages are streamed via a Redis Pub/Sub broker, and recent task data are buffered in Pandas DataFrame, which implements the \texttt{In-memory context}. New user-defined query \texttt{Guidelines} and the \texttt{Dynamic Dataflow Schema} are maintained as lightweight in-memory structures incrementally updated as new messages arrive.

Agent tools and prompt-based interactions are implemented as modular MCP server endpoints. LLM responses are generated using open-source and proprietary models, including \textbf{LLaMA 3} (8B and 70B) hosted on Oak Ridge National Laboratory (ORNL)'s cloud, \textbf{GPT}-4 on Microsoft Azure, and \textbf{Gemini} 2.5 Flash Lite and \textbf{Claude} Opus 4 on Google Cloud Platform. All LLMs had their temperatures set to zero to reduce randomness. 
To evaluate and iteratively refine our agent, we use two complementary use cases. 

\begin{figure*}[!ht]
  \centering
  \includegraphics[width=\textwidth]{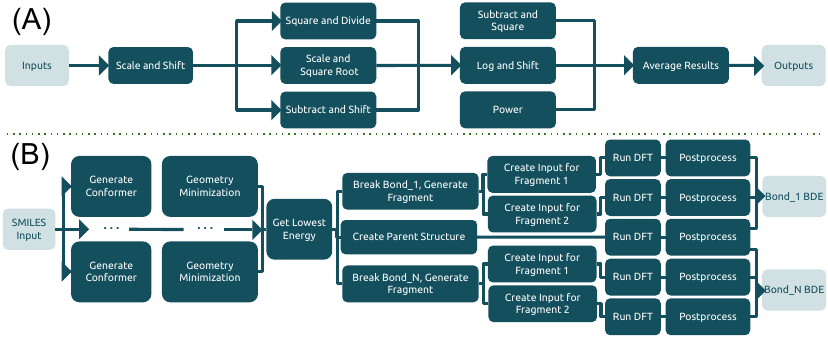}
  \caption{Use case workflows: (A) Synthetic math workflow; (B) Real computational chemistry workflow for Bond Dissociation Energy (BDE) analysis using Density Functional Theory (DFT). Ellipses (`...') indicate repeated steps for multiple conformers; `N' represents the number of bonds in the target molecule.}
  \label{fig:use-cases}
\end{figure*}

\vspace{3pt}
\noindent 
\textbf{Use Case 1: synthetic workflow.} 
The is a synthetic, fast-executing workflow specifically designed to support rapid agent prototyping and prompt fine-tuning (Figure~\ref{fig:use-cases}-A). This workflow is lightweight, free of external dependencies, and enables fast iteration over prompt and system-level changes. Its deterministic behavior allows us to tightly control schema complexity, scale the number of workflow instances, and assess model accuracy across a wide variety of query classes. As a result, it has been instrumental in bootstrapping and stress-testing early agent capabilities. It consists of a small set of chained mathematical transformations forming a fan-out/fan-in structure that exercises both data dependency tracking and semantic reasoning over intermediate states.

\vspace{3pt}
\noindent 
\textbf{Use Case 2: computational chemistry workflow.} 
The chemistry workflow (Figure~\ref{fig:use-cases}-B) prepares, executes, and analyzes density functional theory (DFT) calculations to evaluate molecular energetics and related chemical characteristics. These analyses are critical for understanding complex processes involved in combustion, atmospheric transformations, and redox reactions prominent in biochemical and environmental systems~\cite{Etz-Atmospheric}. This workflow is inspired by previous work using machine learning to predict the bond dissociation enthalpies (BDEs) in an organic molecule~\cite{StJohn_BDE}. It provides a foundational analysis for investigating chemical reactivity. 

The workflow takes a SMILES string as input and orchestrates tools to find the lowest-energy conformation, fragment the molecule, and run geometry optimizations, energy calculations, and vibrational analyses on both parent and fragment structures. It prepares DFT inputs, submits HPC jobs, and generates thermodynamic properties and BDEs for each bond. Compared to the synthetic workflow, it has a more complex dataflow schema with nested structures and chemistry-specific semantics, making it ideal for testing our schema-driven approach.  Both workflows are instrumented with Flowcept decorators for runtime provenance capture.

\vspace{3pt}
\noindent 
\textbf{Reproducibility.} The software used in this work is open source under the MIT license. Flowcept (v0.9), the provenance agent~\cite{flowcept}, synthetic workflow, data, analysis code, query set, LLM prompts~\cite{flowcept_agent,flowcept}, and the chemistry workflow~\cite{compchem} are all available on GitHub.

\input{sections/llm_evaluation}

\subsection{Live Interaction with a Chemistry Workflow}
\label{sec:live-interaction}

\begin{figure}[htbp]
  \centering
  \includegraphics[width=\columnwidth]{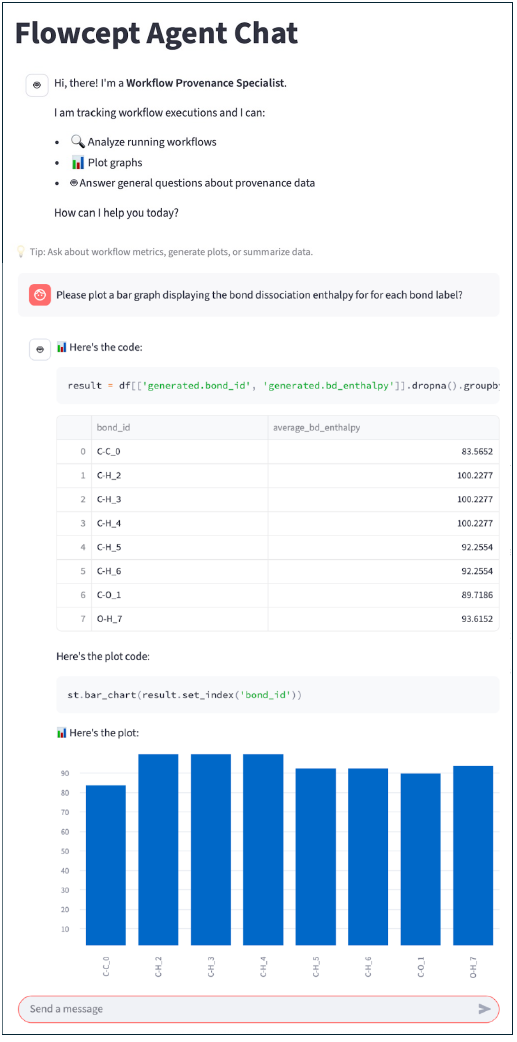}
  \caption{Live interaction with the chemistry workflow. The user interacts in natural language and receives responses, including plots, tabular results, and summarized text.}
  \label{fig:live-interaction}
\end{figure}

To showcase the capabilities of the provenance agent in a real scientific scenario, we conducted a live demonstration that included executing the chemistry workflow on the Oak Ridge Leadership Computing Facility's Frontier supercomputer.
We select ethanol as a simple yet structurally diverse molecule, containing multiple bond types. During the experiment, a scientist interacts with the agent via a web interface, issuing natural language queries at runtime. Responses are generated using GPT-4, with similar results observed from LLaMA 3 70B and Claude Opus 4. LLaMA 3 8B struggles due to its limited context window, as the workflow's dataflow schema is more complex than in the synthetic one. We do not run with Gemini 2.5 Flash Lite due to high response variability (see Section~\ref{sec:agent_eval}). The agent interprets queries, retrieves context, and responds within seconds with tables, plots, or summaries, supporting real-time monitoring and hypothesis validation. Figure~\ref{fig:live-interaction} shows a live interaction screenshot, with more examples available on GitHub~\cite{flowcept_agent}. We now present the queries and discuss their outputs.

\vspace{3pt}\noindent
\textbf{Q1: Which bond has the highest dissociation free energy?}

\noindent \textit{Result: correct.} The agent correctly identified the bond with the highest dissociation free energy and inferred the correct unit (kcal/mol) despite no unit being provided. Among several energy values produced by the BDE workflow (enthalpy, free energy, and electronic energy), the agent chose the correct one.

\vspace{3pt}\noindent
\textbf{Q2: What functional was used for the calculations?}

\noindent \textit{Result: correct; the summary is perfect, but the tabular result could be more concise.} 
Although a concise summary properly presented the correct DFT functional used in the analysis (`B3LYP'), the tabular result displayed repeated values across all calculations. This display could get messy for larger chemicals with many connections.

\vspace{3pt}\noindent
\textbf{Q3: What is the lowest energy bond enthalpy?}

\noindent \textit{Result: correct, but with unit error.} The agent correctly identified the BDE value but used the wrong unit (kJ/mol) and omitted the bond ID (expected: C–C). While the core logic was sound, the lack of unit and bond ID in the query forced the agent to infer them, which can lead to misleading interpretations in chemical reactivity analysis.

\vspace{3pt}\noindent
\textbf{Q4: What is the number of atoms in this molecule?}

\noindent \textit{Result: correct, but ambiguous.} The agent identified the correct atom counts across parent and fragment molecules and presented all values in a table. This is important for validating molecule fragmentation and ensuring that chemical structures remain consistent. However, the response did not clearly associate atom counts with the specific molecule labels, which leads to reduced interpretability.

\vspace{3pt}\noindent
\textbf{Q5: What is the number of atoms in the parent molecule?}

\noindent \textit{Result: incorrect.} The agent incorrectly summed the atom counts from all molecules, returning a total of 81 rather than the number for just the parent structure (should be 9 atoms). 

\vspace{3pt}\noindent
\textbf{Q6: What are the multiplicity and charge of the parent?}

\noindent \textit{Result: correct.} The agent produced accurate information and even enriched it with relevant chemical terminology, such as ``singlet state'' and ``neutral charge'', without being informed about these terms. This highlights the potential of improving the agent's domain knowledge related to chemical semantics.

\vspace{3pt}\noindent \textbf{Q7: Plot a bar graph displaying the bond dissociation enthalpy for each bond label.}

\noindent  \textit{Result: correct.} The agent generated a plot (Figure~\ref{fig:live-interaction}) accurately showing BDE by bond label. This visualization helps chemists compare bond stability and reactivity, and spot patterns or outliers. 

\vspace{3pt}\noindent
\textbf{Q8: For this molecule, please plot a bar graph displaying the bond dissociation enthalpy with averaged C–H values.}

\noindent \textit{Result: incorrect.} The agent failed to group C–H bonds and compute an average before plotting. While this is a challenging task for the agent, it is less relevant chemically, as individual bond values are more informative. However, supporting such custom visualizations is useful for comparative analysis across bond types.

\vspace{3pt}\noindent
\textbf{Q9: What is the average bond dissociation enthalpy for the bond labels that contain `C-H'?}

\noindent \textit{Result: correct.} Despite the error in Q8, the agent successfully calculated the average BDE value for the five C-H bonds, suggesting that the plot logic, in this implementation, still needs to be improved.

\vspace{3pt}\noindent
\textbf{Q10: What is the multiplicity and charge of any fragment?}

\noindent \textit{Result: correct.} Similar to Q6, the model correctly retrieved the relevant charge and multiplicity data, which is crucial for validating and representing the correct electronic state of the fragment structures. Unlike in Q6, the agent did not include key terms in the summary.

\subsection{Evaluation Summary, Lessons Learned, and Future Work}
\label{sec:discussion}

Our work demonstrates that LLM-powered agents can enable effective interaction with complex, large-scale workflow provenance data when designed with the right abstractions and evaluation methodology. Combining a modular design, flexible provenance capture, and RAG methods powered by a dynamic dataflow schema, we built a lightweight, extensible, and effective AI agent.

\vspace{3pt}
\noindent\textbf{Key Findings.}
The provenance agent that performed well on a simple synthetic workflow also generalized effectively to a more complex real-world use case without requiring additional domain-specific prompt engineering. Originally prototyped with a lightweight mathematical workflow, the agent adapted successfully to a computational chemistry workflow on the Frontier supercomputer, where it either correctly or partially correctly answered over 80\% of queries. These included inferring domain-specific concepts, despite never being explicitly informed about them. This demonstrates both the agent's strong reasoning capabilities and the generality of the lightweight design. In addition to these two workflows, we are already using the agent in a third workflow in the additive manufacturing (metal 3D printing) domain~\cite{souza2025provagent}.

The dynamic dataflow schema was key in this generalization, enabling the agent to track workflow structure and key parameter semantics incrementally, while avoiding context window overflows even in an HPC environment. Because the agent operates on metadata rather than task provenance data, its LLM performance is independent of the number of workflow tasks, volume of processed data, or volume of captured provenance, depending instead on the workflow complexity alone, i.e., the number and diversity of activities and their input and output fields. This trade-off allows for lightweight operation in large workflows. 
These results highlight the approach's potential as a low-barrier, interactive tool to explore and visualize provenance data without requiring structured query languages, custom code, or dealing with the complexities of setting up their own data analysis environments. We believe that this work has the potential to accelerate scientific discovery by reducing the time-to-insights in complex ECH workflows.

\vspace{3pt}
\noindent\textbf{Architecture Features.}
The reference architecture separates concerns for modularity, extensibility, and lightweight deployment. By isolating provenance capture, stream ingestion, dataflow schema construction, prompt generation, and LLM interaction, it enables easy tool integration and scaling. The design works with any provenance system with a queryable backend and supports interactive use across multiple concurrent and agentic workflows.

\vspace{3pt}
\noindent\textbf{Lessons from Prompting and Evaluation.}
We assessed how each contextual component (i.e., query guidelines, dynamic dataflow schema) contributed to individual query classes, understanding which component we should invest further in to improve the responses' quality.  OLAP queries proved the hardest across all LLMs.
The query guidelines and few-shot examples helped reduce syntax errors, while the dataflow schema and domain-specific examples enabled better semantic alignment. 

Our goal was not to benchmark specific LLMs, but to evaluate a metadata, query-driven approach that remains robust as models change. Across the tested models, the same pattern holds: query guidelines, the dynamic dataflow schema, and domain values drive most gains, while having metadata as input and queries as output keep token usage bounded. This design should maintain high scores as newer LLMs appear, without altering our proposed architecture and approach. 
Even as LLMs improve and context windows grow with newer LLMs, their invocations will not ingest a whole provenance database, which may reach gigabytes or terabytes in large runs. Thus, our approach will remain necessary for future LLM versions, scaling by depending on the dataflow schema and queries instead of prompting the entire database.

We also observed that LLM response times, including both processing and cloud access latency, remained within acceptable interactive thresholds (\textasciitilde2s). While this met our goals for interactive data analysis via accurate agent responses, future work could investigate whether specific query classes or contextual components impact latency, potentially revealing strategies to further optimize LLM interactivity with provenance databases.

Our initial system used a static set of query guidelines, which we iteratively refined during early development with the synthetic workflow. As we tested the agent across diverse query classes, we updated these guidelines manually until performance was satisfactory. Applying the agent to the computational chemistry workflow revealed new edge cases, reinforcing that, as  humans, we cannot anticipate all possible scenarios. This led us to redesign the architecture to support dynamic, user-defined query guidelines.

The current GUI displays the code generated and executed on the in-memory DataFrame (Figure~\ref{fig:live-interaction}), including any runtime errors. This allows users to issue corrections or run revised DataFrame code manually, directly within the GUI. Although not ideal, this has proven useful for debugging and improving transparency.
In practice, users or developers can use this mechanism to generalize small fixes into reusable user-defined guidelines, closing the loop between user feedback and prompt adaptation. In the future, we envision replacing this manual flow with a feedback-driven ``auto-fixer'' agent specialized in diagnosing query failures, proposing corrected versions, and automatically suggesting new guidelines.

Evaluation using LLM-as-a-judge was scalable and flexible, enabling us to assess query correctness without exhaustively encoding and curating individual rules. While human supervision remained necessary to validate fairness and scoring consistency, this is an improvement over rigid rule-based evaluation. 
Using multiple judges like GPT and Claude improved reliability: despite minor score differences, their consistent patterns mitigated individual bias. This highlights the value of combining LLM-based scoring with simple guidelines and consensus for more trustworthy results.

\vspace{3pt}
\noindent\textbf{Limitations and Open Challenges.}
Our experiments focused on online queries over the in-memory context to support interactive monitoring of workflows in the ECH continuum. While the architecture supports offline querying, enabling deep graph traversals (e.g., for multi-hop causal analysis) over persistent provenance databases will require significant additional work, as such queries go beyond what DataFrames can easily represent.

Despite the near-perfect overall results from models like GPT and Claude, no single LLM excelled across all query classes, highlighting the potential for future research on intelligent, adaptive LLM routing based on query class.
Another open challenge is the semantic quality of workflow code. Since our approach does not require users to provide semantic schemas or annotations beforehand, and it relies on explicit code instrumentation or implicit data observability, the agent's performance depends on how intentional the user is when writing their code with meaningful variable and function names. While this flexibility encourages adoption, it introduces a dependency on code quality, highlighting the need for future research on inferring semantics from poorly descriptive code.

\vspace{3pt}
\noindent\textbf{Scalability and Future Extensions.}
While we did not benchmark extreme-scale workloads, our metadata-driven design keeps LLM interaction independent of data volume by only accessing schemas and query guidelines, allowing lightweight deployment even for large HPC jobs. Queried monitoring data remain in an in-memory buffer (currently a Pandas DataFrame), which is efficient when data fit in memory~\cite{Petersohn2020TowardsSD}. Migrating to higher-performance options like Polars would be straightforward and could benefit extreme-scale workflows that generate massive amounts of provenance data.


Overall, our findings show that LLM-based agents can effectively interact with workflow provenance databases when supported by a modular architecture, dynamic schema, and well-crafted prompting strategy. Our approach yielded promising results in a live interaction with a chemistry workflow on the Frontier supercomputer and lays the groundwork for future research in intelligent workflow analysis, steering, and reproducibility. 
Nevertheless, achieving truly natural and seamless interaction with complex workflow provenance databases remains an open challenge. We view our agentic approach as one component in a broader toolchain. Particularly, in very complex ECH workflows, traditional data analysis pipelines may still be necessary to handle edge cases. Continued research in interactive provenance systems will be essential to accelerate the data-to-insights process and, ultimately, scientific discovery.

%% file: sections/llm_evaluation.tex
\subsection{LLM Agent Response Analysis}
\label{sec:agent_eval}

The objective of this section is to evaluate and iteratively improve the LLM-powered provenance agent by applying our methodology (Section~\ref{fig:methodology}). Rather than a static assessment, our evaluation process served as a feedback loop, helping us refine our prompt engineering strategies and RAG pipeline. Each experiment run informed the next, allowing us to incrementally improve agent performance while also demonstrating the practicality and effectiveness of the methodology in guiding the development of this provenance agent.

The methodology starts by defining a set of 20 natural language queries, each labeled with a query class (i.e., leaves in Figure~\ref{fig:query_classes}) and a corresponding expected DataFrame code snippet. The queries, evenly split between OLAP and OLTP, were manually curated. Since some involve multiple provenance types (e.g., telemetry and scheduling), the data type totals exceed 20. Table~\ref{tab:query_distribution} shows the distribution of queries across the data types. 
While Flowcept supports database systems for persisting provenance data, this evaluation focuses on online retrospective queries over recent or active workflow runs using an in-memory context, aligning with our target use case: interactive chats with provenance data during workflow execution.

\input{tables/query_counts}


We analyze how contextual information from prompt engineering and RAG affects LLM performance. The context has two components: prompt elements and RAG-derived schema data. Prompt elements include the agent role (e.g., ``Your job is to query a live DataFrame buffer"), a DataFrame description (e.g., ``Each row represents a task execution"), output formatting (e.g., ``Return a single DataFrame query"), few-shot examples (natural language + DataFrame code pairs), and query guidelines (e.g., ``When filtering time ranges, use the field \texttt{started\_at}"). The RAG input provides a dynamic dataflow schema with available field names, example domain values, and inferred types or shapes for arrays. In each experiment we toggle components to isolate their contribution to the response score; this layered design with our query taxonomy enables fine-grained evaluation across query classes and pinpoints where accuracy and generalization can improve. Table~\ref{tab:prompt_rag_configs} summarizes the incremental configurations, from zero-shot (user query only) to full context.

\input{tables/context_table}

To evaluate query accuracy, we use an LLM-as-a-judge approach with a tailored prompt that instructs the model to act as an expert evaluator. It compares a gold standard query, written by a human, with the agent-generated query, both addressing the same user input. The judge has access to the same context as the provenance agent.
The prompt emphasizes functional equivalence over syntactic similarity, encouraging high scores if the generated query achieves the same analytical goals, even with structural differences. It also outlines edge cases, such as invalid column references or incorrect logic, and defines a scoring scale from 0.0 to 1.0 based on how well the generated code satisfies the user's intent. This strategy enables scalable, consistent, and nuanced evaluation of agent responses with less human supervision as compared to a rule-based approach, where we would need a human to carefully define several rules to match the generated query. 
To improve reliability and reduce bias in evaluation, we use two different LLMs as judges: GPT and Claude. Because LLMs can still produce slight variations even with the temperature set to zero, each query is executed three times, and we get the median results per query. We run the synthetic workflow with 100 input configurations; results remain consistent across runs with as few as 1 and as many as $1,000$ inputs, reflecting the metadata- and query-oriented design that is independent of provenance data volume.

\begin{figure}[!ht]
  \centering
   \includegraphics[width=0.8\columnwidth]{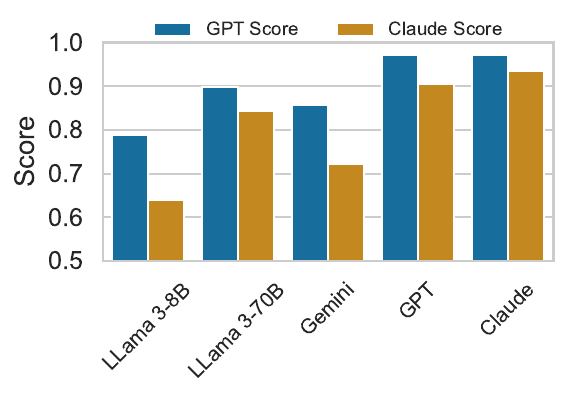}
  \caption{Scores assigned by two different judges.}
  \label{fig:plot:judge_comparison}
\end{figure}

\vspace{3pt}
\noindent\textbf{Comparing the Two Judges}.
Figure~\ref{fig:plot:judge_comparison} compares the average of median scores assigned by GPT and Claude as judges across five evaluated LLMs. Overall, GPT judge consistently scores responses higher than Claude, with the largest differences observed for LLaMA 3–8B and Gemini. 
Although the ranking trend remains consistent across judges, absolute scores differ, reflecting individual scoring tendencies. Each judge appears to slightly favor its own model: GPT rated GPT at 0.972 and Claude at 0.970 (in practice a tie within expected error margins), while Claude rated Claude at 0.94 and GPT at 0.91 (a more noticeable difference). This bias emerges despite the double-blind setup, where judges were not told which model they were evaluating. These results highlight the need for multiple judges to balance individual biases. Still, the most important finding is the strong agreement pattern across judges, which reinforces the reliability of the overall evaluation.

\begin{figure*}[!ht]
  \centering
  \includegraphics[width=\textwidth]{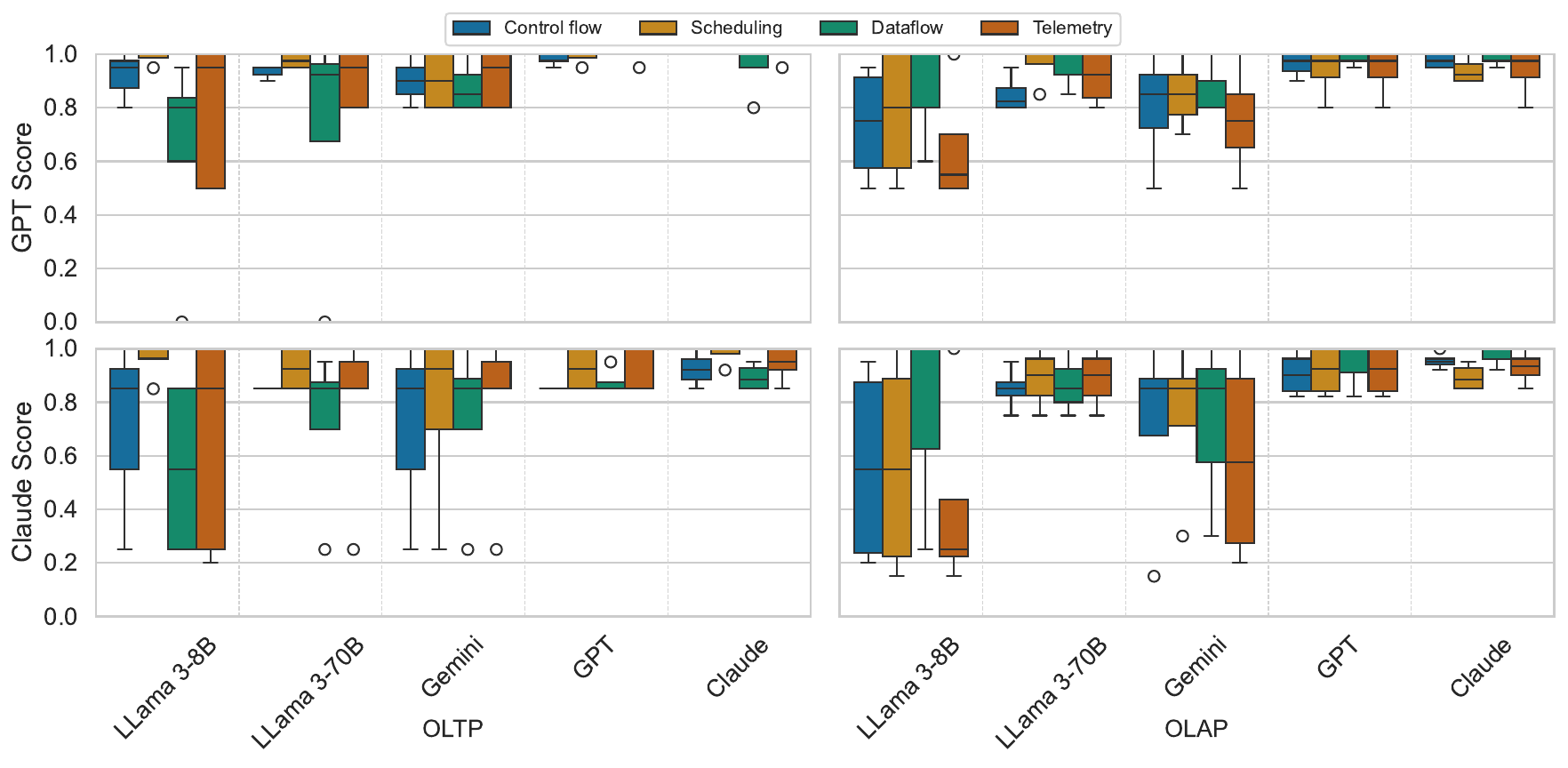}    
  \caption{Different LLMs' performance in different query classes.}
  \label{fig:plot:query-classes}
\end{figure*}

\vspace{3pt}
\noindent\textbf{Comparing LLMs in Different Query Classes}.
Figure~\ref{fig:plot:query-classes} shows per-class model scores under the full context configuration. Each subplot presents boxplots of median scores by model and data type, separated by workload type (OLAP or OLTP) and evaluated independently by GPT and Claude judges. This enables analysis of how each LLM generalizes across diverse query classes and provenance data types under a consistent prompt and RAG setup.

Across both judges, OLTP queries tend to yield higher scores and less variation, indicating that most models can handle transactional-style questions well. In contrast, OLAP queries show greater dispersion and more frequent low scores, reflecting their higher complexity and need for logical reasoning. For data types, Scheduling and Telemetry queries generally receive higher scores, especially under OLTP. 
The LLM-as-a-judge feedback shows that Dataflow and Control Flow are more error-prone due to their need to interpret graph-like relationships and nested logic.

GPT and Claude consistently outperform other models across all data types and workloads, with only subtle differences in how each judge scores them. Averaged over all queries, GPT and Claude receive nearly identical scores from the GPT judge, while Claude slightly outperforms GPT by 0.03 according to the Claude judge. In OLAP workloads, both judges assign similarly high scores to the two models, with slight variations: Claude favors GPT in Control Flow and Scheduling, while GPT favors Claude in Telemetry. In OLTP, both judges give perfect or near-perfect scores. These results reflect strong agreement between the models and judges, but with each judge showing a mild preference for its own outputs.

The judges' feedback also shows that LLaMA 3–8B often hallucinated non-existing fields like \texttt{node} or \texttt{execution\_id} and ignored guidelines. LLaMA 3–70B struggled with \texttt{group by} logic or time comparisons. Gemini's performance has the greatest variability, especially on OLAP Telemetry. Claude's and GPT-4's errors typically involved logic misinterpretations (e.g., using \texttt{.min()} on IDs instead of timestamps). These findings suggest that no single model performs best across all workloads and data types, motivating future research on dynamic LLM routing based on query classes.

\vspace{3pt}
\noindent\textbf{Evaluating Impact of Contextual Information Components against Performance and Token Consumption}.
Considering previous findings with full context, we now analyze how individual prompts and RAG components affect performance and token usage. This helps prioritize components with the highest gains and identify those needing refinement. We evaluate six cumulative configurations, from the baseline to full context (Table~\ref{tab:prompt_rag_configs}), using the GPT model and judge for their consistently good performance. Zero-shot setups were excluded due to consistently poor scores across all models, underscoring the importance of prompt tuning and schema- and guideline-informed RAG.

\begin{figure}
  \centering
  \includegraphics[width=\columnwidth]{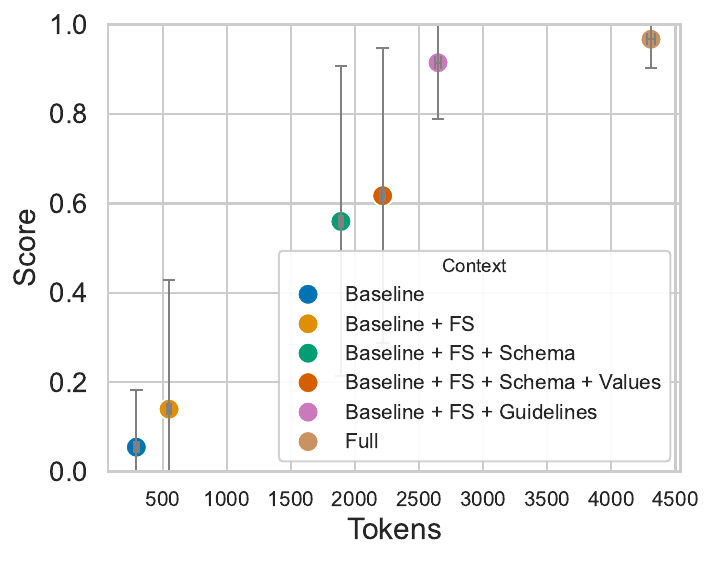}
  \caption{Impact of contextual components on performance and token consumption. See Table \ref{tab:prompt_rag_configs} for labels.}
  \label{fig:plot:contexts-tokens}
\end{figure}

Figure~\ref{fig:plot:contexts-tokens} shows the average of median performance and total token usage (input + output) across six cumulative configurations. Each point represents the mean of median scores and token counts per query, with standard deviation as error bars. As context components are added from Baseline to Full, average scores rise from 0.06 to 0.97, with the largest jump between FS + Schema (0.56) and FS + Guidelines (0.92). Token usage grows from 293 to over 4,300, approaching the limits of smaller models like LLaMA 3–8B (\textasciitilde8k), while remaining well below the limits of larger models such as GPT-4o (128k). Schema descriptions and domain values are the most token-expensive, while few-shot examples and query guidelines yield significant performance gains with small token overhead.

\begin{figure}[!ht]
  \centering
  \includegraphics[width=\columnwidth]{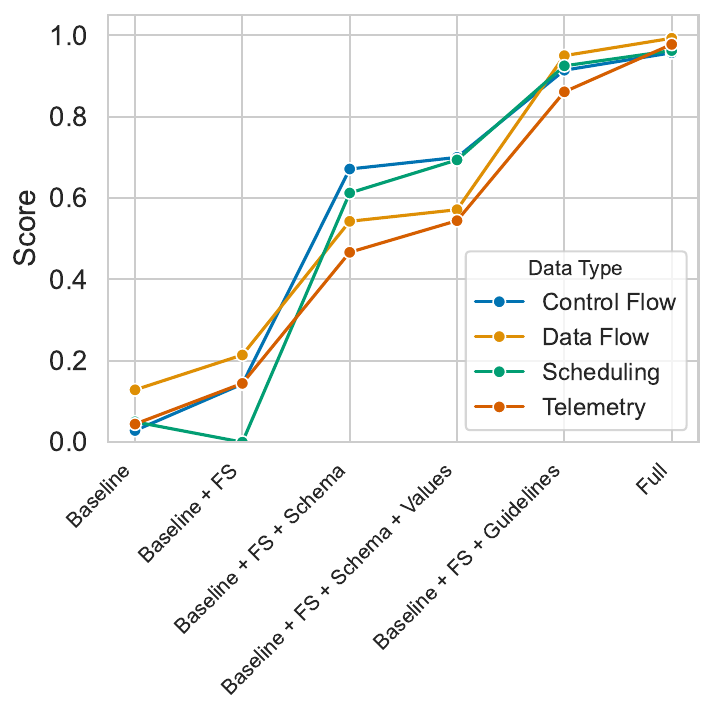}
  \caption{Impact of contextual components on performance and data types.  See Table \ref{tab:prompt_rag_configs} for labels.}
  \label{fig:plot:contexts-data_types}
\end{figure}

\vspace{3pt}
\noindent\textbf{Evaluating Impact of Contextual Information Components on Different Data Types.}
To understand the influence of prompt and RAG contextual components on different data types, we analyze GPT performance (judged by GPT) across the different data types. Figure~\ref{fig:plot:contexts-data_types} shows that all data types benefit from richer context. For example, Control Flow scores improve from 0.70 to 0.91 with guidelines, and Dataflow jumps from 0.57 to 0.95, highlighting guidelines' critical role in helping the model interpret intent and select correct fields. The dataflow schema and example domain values also boost performance for schema-dependent types, such as Dataflow. In contrast, Scheduling and Telemetry show more stable improvements across configurations, starting lower (e.g., Telemetry 0.04) and gradually reaching 0.96–0.98 in the Full setting, showing that they benefit more uniformly from general context components such as task-level examples and guidelines.

\vspace{3pt}
\noindent\textbf{Response times.} We evaluate LLM response times based on the duration of HTTP API calls to the cloud-hosted LLM service, using the mean of per-query median latencies. Results show stable performance across both OLAP and OLTP workloads and consistency across data types. Even with full-context prompts, all models stay within interactive latency bounds (\textasciitilde2s), confirming the agent's suitability for live workflow interaction.


\vspace{3pt}
\noindent\textbf{Summary.}
GPT-4 and Claude Opus 4 consistently achieve near-perfect scores, while LLaMA and Gemini show greater variability and lower performance. OLAP queries involving graph-like data remain the most challenging for all LLMs. Among contextual components, query guidelines provide the greatest performance boost with lower token cost. No single LLM is the best across all query types, revealing the need for adaptive LLM routing. While GPT and Claude judges differ slightly in absolute scores, their consistent agreement trends reinforce confidence in evaluation reliability.

%% file: tables/query_counts.tex

\begin{table}[ht]
\scriptsize
\centering
\setlength{\tabcolsep}{4pt} 
\caption{Distribution of queries by data type and workload.}
\label{tab:query_distribution}
\arrayrulecolor{gray!50}
\resizebox{0.85\linewidth}{!}{%
\begin{tabular}{@{}p{0.30\linewidth}
                >{\centering\arraybackslash}p{0.15\linewidth}
                >{\centering\arraybackslash}p{0.15\linewidth}
                >{\centering\arraybackslash}p{0.15\linewidth}@{}}
\toprule
\textbf{Data Type} & \textbf{OLAP} & \textbf{OLTP} & \textbf{Total} \\
\midrule
Control Flow & 4 & 3 & 7 \\
\midrule
Dataflow & 3 & 4 & 7 \\
\midrule
Scheduling & 3 & 5 & 8 \\
\midrule
Telemetry & 4 & 5 & 9 \\
\bottomrule
\end{tabular}
}
\arrayrulecolor{black}
\end{table}

%% file: tables/context_table.tex
\begin{table}[ht]
\scriptsize
\centering
\caption{Prompt + RAG configurations used for evaluation.}
\label{tab:prompt_rag_configs}
\arrayrulecolor{gray!50}
\begin{tabular}{@{}p{0.65\linewidth}p{0.25\linewidth}@{}}
\toprule
\textbf{Context (Prompt+RAG strategy)} & \textbf{Label} \\
\midrule
Zero-shot & Nothing \\
\midrule
Role + Job + DataFrame format + Output Formatting & Baseline \\
\midrule
Baseline + Few shot & Baseline+FS \\
\midrule
Baseline + Few Shot + Dynamic Dataflow Schema & Baseline+FS+Schema \\
\midrule
Baseline + Few Shot + Dynamic Dataflow Schema + Domain Values & Baseline+FS+Schema+ Values \\
\midrule
Baseline + Few Shot + Query Guidelines & Baseline+FS+Guidelines \\
\midrule
Baseline + Few Shot + Dynamic Dataflow Schema + Domain Values + Query Guidelines & Full \\
\bottomrule
\end{tabular}
\arrayrulecolor{black} 
\end{table}

%% file: sections/conclusion.tex
\section{Conclusions}

Our work demonstrates that LLM-powered agents, when guided by structured schema representations, dynamic prompting strategies, and a modular system architecture, can enable effective, near real-time interaction with complex workflow provenance data across the ECH continuum. By decoupling LLM interaction from raw data volume and focusing on metadata, our approach remains lightweight and scalable, even for large HPC workflows. The agent generalized well from synthetic to real-world settings, achieving high accuracy without domain-specific tuning. These results suggest that interactive, provenance-aware AI agents can significantly reduce the effort required for exploratory data analysis, anomaly diagnosis, and monitoring, closing the gap between scientists and their data. We believe this work lays the foundation for a new class of intelligent provenance agents while highlighting open challenges such as dynamic semantic enrichment of schemas, feedback-driven autonomous prompt tuning, and graph-based LLM querying.

%% file: sections/ack.tex
\medskip
{\footnotesize
\noindent \textbf{Acknowledgments.}
ChatGPT-4o was used to help polish writing, improve conciseness, and check grammar across the sections of the paper. This research used resources of the Oak Ridge Leadership Computing Facility at Oak Ridge National Laboratory, supported by the U.S. Department of Energy Office of Science under Contract No. DE-AC05-00OR22725. It was also supported in part by an appointment to the Education Collaboration at ORNL (ECO) Program, sponsored by the DOE and administered by the Oak Ridge Institute for Science and Education, and the U.S. Department of Energy Office of Science under Contract No. DE-AC02-06CH11357.
}